# Energy Spectra of Abundant Nuclei of Primary Cosmic Rays from the Data of ATIC-2 Experiment: Final Results

A. D. Panov[a], J. H. Adams Jr.[b], H. S. Ahn[c], G. L. Bashinzhagyan[a], J. W. Watts[b], J. P. Wefel[d], J. Wu[c], O. Ganel[c], T. G. Guzik[d], V. I. Zatsepin[a], I. Isbert[d], K. C. Kim[c], M. Christl[b], E. N. Kouznetsov[a], M. I. Panasyuk[a], E. S. Seo[c], N. V. Sokolskaya[a], J. Chang[f, g], W. K. H. Schmidt[g], and A. R. Fazely[e]

[a] Skobel'tsyn Institute of Nuclear Physics, Moscow State University, Moscow, Russia
e-mail: panov@dec1.sinp.msu.ru

[b] Marshall Space Flight Center, United States

[c] University of Maryland, United States

[d] Southeastern Louisiana University, United States

[e] Southern University, Baton Rouge, Louisiana, United States

[f] Purple Mountain Observatory, China

[g] Max Planck Institute, Germany

**Abstract**—The final results of processing the data from the balloon-born experiment ATIC-2 (Antarctica, 2002–2003) for the energy spectra of protons and He, C, O, Ne, Mg, Si, and Fe nuclei, the spectrum of all particles, and the mean logarithm of atomic weight of primary cosmic rays as a function of energy are presented. The final results are based on improvement of the methods used earlier, in particular, considerably increased resolution of the charge spectrum. The preliminary conclusions on the significant difference in the spectra of protons and helium nuclei (the proton spectrum is steeper) and the non-power character of the spectra of protons and heavier nuclei (flattening of carbon spectrum at energies above 10 TeV) are confirmed. A complex structure of the energy dependence of the mean logarithm of atomic weight is found.

**DOI:** 10.3103/S1062873809050098

## INTRODUCTION

The balloon-born experiment ATIC-2 (Advanced Thin Ionization Calorimeter) is designated for measuring the composition and energy spectra of primary cosmic rays in the energy range from approximately 50 GeV to 100–200 TeV per particle with the element-to-element charge resolution from protons to iron, and the electron spectrum of primary cosmic rays.

The ATIC spectrometer consists of a silicon matrix for determining the charge of the primary particle, a graphite target, scintillator hodoscopes, and a completely active BGO calorimeter. The detailed description of the spectrometer and calibration methods was given in [1], the method for determining the charge of the primary particle, including details related to solution of the problem of albedo particles using the highly segmented matrix of silicon detectors, was described in [2, 3]. The calorimeter is thin, i.e., it detects only some part of the primary particle energy. The methods for reconstructing the energy spectrum of primary particles from the energy release spectrum were described in [4, 5].

The ATIC instrument performed three stratospheric flights about the South Pole: ATIC-1 (December 28, 2000–January 13, 2001), ATIC-2 (December 29, 2002–

January 18, 2003), and ATIC-4 (December 26, 2007–January 15, 2008). The total flight duration was about 50 days. The flight of ATIC-3 in December, 2005, turned out to be unsuccessful because of the balloon shell damage at the takeoff. The ATIC-1 flight was a test one and the ATIC-4 data are now being processed. The preliminary results of processing the ATIC-2 data related to the hadronic component of cosmic rays were published in [5–8]. In this paper, we report the final results of processing the data of this experiment on the energy spectra of abundant nuclei (protons, He, C, O, Ne, Mg, Si, and Fe nuclei), for the spectrum of all particles, and for the energy dependence of the mean logarithm of atomic weight of primary cosmic rays. A number of new methods were used to obtain these results. The presented results are final, since all basic known methodical effects were taken into account in processing. Further improvement of methods cannot result in noticeable revision of results.

## NEW METHODS AND RESULTS

### Proton and Helium Spectra

In previous papers [5, 6, 8], the overlap of charge lines of protons and helium nuclei was taken into account approximately using the correction coefficients. In this study, were used the FLUKA code for primary protons and helium nuclei with different initial energies to perform quantitative simulation of the instrumental shape of charge lines. The problem of reconstructing the initial intensities of overlapping proton and helium lines was solved accurately taking into account the obtained functions. The refined processing data are shown in Fig. 1. The refined results differ only slightly from the more approximate preliminary data [5, 8]. These results confirm the conclusion on different average slopes of the proton and helium spectra: the helium spectrum is on average flatter. Figure 1 shows, along with the ATIC-2 data, the results of the AMS [9, 10], CAPRICE-98 [11], and BESS-TeV [12] experiments.

*Spectra of Abundant Even Nuclei from Carbon to Iron*

The following methodical effects were taken into account upon obtaining spectra of abundant nuclei beginning from carbon.

(1) Backgrounds related to the high-energy tails of charge distributions for proton and helium lines, which result from either fluctuations of ionization losses in silicon detectors (Landau distribution tails) or the influence of inverse currents and nuclear interactions of particles in the charge unit were taken into account. Due to these high-energy tails, a primary proton or helium nucleus can simulate the arrival of a heavier nucleus, which can distort noticeably the values of fluxes of nuclei heavier than helium.

(2) The FLUKA code was used to calculate the instrumental shapes of charge lines for each nucleus of interest and the cross backgrounds of different nuclei (due to the interactions with silicon matrix design elements, a nucleus can be falsely recorded as another nucleus with either higher or lower charge). The problem of taking into account the line overlap was solved accurately using this information (it is reduced to solution of a system of linear equations).

(3) The logarithmic growth of the ionization power of particles (shift of charge lines) with increasing particle energy. In fact, were used the connection of the position of charge lines with the energy release in the calorimeter rather than with the initial particle energy, and this dependence was approximated by a linear function of the energy logarithm with the slope $\alpha = 1.2\%$ per decade of energy release. It was shown that the coefficient $\alpha$ depends weakly on the particle charge (as was expected).

(4) The upper layer of the scintillator hodoscope was used as the additional charge detector; this approach allowed us to obtain the charge spectrum with a much better resolution (the line full width at half-maximum for the carbon line was on average $0.77e$ for the silicon matrix and $0.59e$ for the combination of silicon matrix and scintillator hodoscope) at the expense

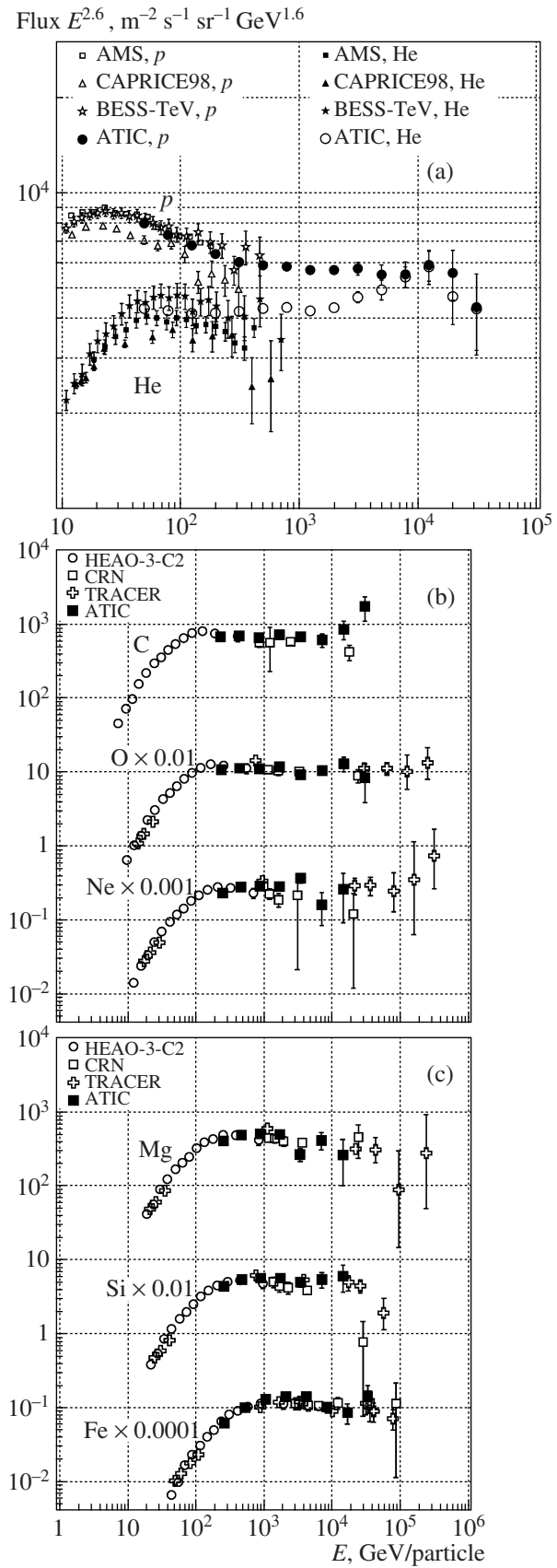

**Fig. 1.** Spectra of protons, helium, and other abundant nuclei.

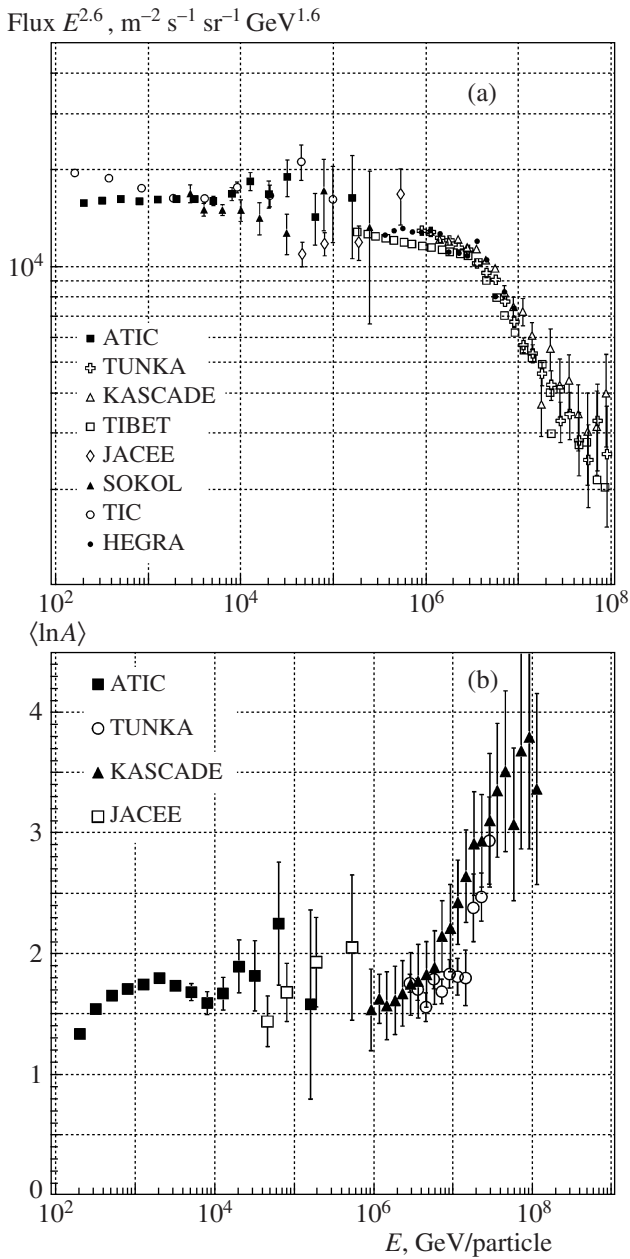

Flux $E^{2.6}$, $m^{-2}$ $s^{-1}$ $sr^{-1}$ $GeV^{1.6}$

(a)

- ■ ATIC
- ◇ TUNKA
- ▲ KASCADE
- □ TIBET
- ◇ JACEE
- ◇ SOKOL
- ○ TIC
- • HEGRA

$\langle \ln A \rangle$

(b)

- ■ ATIC
- ○ TUNKA
- ▲ KASCADE
- □ JACEE

$E$, GeV/particle

**Fig. 2.** (a) Spectrum of all particles and (b) the mean logarithm of atomic weight $\langle \ln A \rangle$ as a function of energy.

liminary data [5–7] by a more pronounced flattening of the carbon spectrum at energies higher than 10 TeV.

*Spectrum of All Particles and the Energy Dependence of the Mean Logarithm of Atomic Weight*

Previously [5], the spectrum of all particles was obtained by summing the spectra of separate classes of nuclei which were reconstructed from the energy release spectra using different methods: the proton and helium spectra were determined by deconvolution with Tikhonov's regularization [4], while the spectra of heavier nuclei were treated using the differential scaling (shift) method [5]. As a result, we could obtain the spectrum only to energies of about 50 TeV. In this study, all partial spectra were obtained using the differential shift method, and the spectrum of all particles was determined to energies of about 150 TeV (Fig. 2). The energy dependence of the mean logarithm of the atomic weight of hadronic component of cosmic rays for the same energy range was also determined. The spectrum of all particles was compared with the data of the Tunka [16], KASCADE [17], TIBET [18], JACEE [19], Sokol [20], TIC [21], HEGRA [22] experiments. The energy dependence of the logarithm of atomic weight was compared with the results of the KASCADE [17], JACEE [19], and Tunka (V. V. Prosin, private communication, 2008) experiments. The data on $\langle \ln A \rangle$ of the ATIC experiment and the data of other experiments point to the complex structure of the energy dependence of the cosmic ray mass composition.

ACKNOWLEDGMENTS

This work was supported by the Russian Foundation for Basic Research, project nos. 02-02-16545 and 05-02-16222, and NASA grants NNG04WC12G, NNG04WC10G, and NNG04WC06G.

D. Chang acknowledges the support of the Ministry of Science and Technology of China (grant no. 2002CB713905).

of some reduction of statistics (by 40%). This made it possible to perform the entire processing in two different ways (for spectra with high statistics and low resolution and for spectra with lower statistics and higher resolution) and compare the results. The agreement was the additional substantiation of the correctness of both methods. Figure 1 shows the found C, O, Ne, Mg, Si, and Fe spectra. The ATIC-2 data are compared with the results of the HEAO-3-C2 [13], CRN [14], TRACER [15] experiments. The new results differ from the pre-